\documentclass{pasj00}
\draft

\begin{document}
\SetRunningHead{Fukushige {\it et al.}}{GRAPE-6A}
\Received{}%{yyyy/mm/dd}
\Accepted{}%{yyyy/mm/dd}

\title{GRAPE-6A: A single-card GRAPE-6 for parallel PC-GRAPE cluster
system}

\def\JM#1{{\bf [JM: #1]}}

%%% begin:list of authors
\author{Toshiyuki \textsc{Fukushige}}
\affil{Department of General System Studies, College of Arts and Sciences,\\
University of Tokyo, Tokyo 153-8902, Japan}
\email{fukushig@provence.c.u-tokyo.ac.jp}

\author{Junichiro \textsc{Makino}}%
\affil{Department of Astronomy, School of Science, \\
University of Tokyo,  Tokyo 133-0033, Japan}
\email{makino@astron.s.u-tokyo.ac.jp}

\and
\author{Atsushi \textsc{Kawai}}
\affil{Faculty of Human and Social Studies,\\
Saitama Institute of Technology, Saitama 369-0293, Japan}
\email{kawai@sit.ac.jp}
%%% end:list of authors

%%% Please use the following style in case that sorting by 
%%% affiliation is impossible. 
%
% \author{%
%   D-Firstname \textsc{D-Familyname}\altaffilmark{1}
%   E-Firstname \textsc{E-Familyname}\altaffilmark{1,2}
%   and
%   F-Firstname \textsc{F-Familyname}\altaffilmark{2}}
% \altaffiltext{1}{Address of Institute}
% \email{ddddd@xxx.xxx.xx.xx}
% \email{eeeee@xxx.xxx.xx.xx}
% \altaffiltext{2}{Address of Institute}

%% `\KeyWords{}' always has to be placed before `\maketitle'.
 %Do NOT move this preamble from here!

\maketitle

\begin{abstract}

In this paper, we describe the design and performance of GRAPE-6A, a
special-purpose computer for gravitational many-body simulations.  It
was designed to be used with a PC cluster, in which each node has one
GRAPE-6A. Such configuration is particularly effective in running
parallel tree algorithm.  Though the use of parallel tree algorithm
was possible with the original GRAPE-6 hardware, it was not very
cost-effective since a single GRAPE-6 board was still too fast and too
expensive.  Therefore, we designed GRAPE-6A as a single PCI card to
minimize the reproduction cost and optimize the computing speed.  The
peak performance is 130 Gflops for one GRAPE-6A board and 3.1 Tflops
for our 24 node cluster. We describe the implementation of the tree,
TreePM and individual timestep algorithms on both a single GRAPE-6A
system and GRAPE-6A cluster.  Using the tree algorithm on our 16-node
GRAPE-6A system, we can complete a collisionless simulation with 100
million particles (8000 steps) within 10 days.

\end{abstract}

%%%%%%%%%%%%%%%%%%%%%%%%%%%%%%%%%%%%%%%%%%%%%%%%%%%%%%%%%%%%
\section{Introduction}

Large-scale $N$-body simulations, in which the equations of motion of
$N$ particles are integrated numerically, have been extensively used
in the studies of galaxies and cosmological structure formations. At
present, such simulations are in most cases performed with fast and
approximate algorithms that reduce the calculation cost from $O(N^2)$
of the direct summation. Examples of such algorithms include P$^3$M
method (Efstathiou and Eastwood 1981) and hierarchical tree algorithm
(Barnes and Hut 1986), and their derivatives (Couchman 1991, Xu 1995,
Bagla and Ray 2003). In order to achieve the best performance, these
algorithms have been implemented on massively parallel supercomputers
(e.g. Pearce and Couchman 1997, Springel, Yoshida, White 2001) or Beowulf-type
PC-clusters (e.g. Dubinski et al. 2003, Diemond et al 2004).

An alternative approach to achieve high computational speed is to use a
specialized hardware for gravitational interaction, GRAPE (GRAvity
piPE) (Sugimoto et al. 1991, Makino and Taiji 1998). A GRAPE hardware
has specialized pipelines for gravitational force calculation, which
is the most expensive part of most of $N$-body simulation
algorithms. All other calculations, such as the time integration of
orbits, are performed on a standard PC or workstation (host computer)
connected to GRAPE. Compared to the calculation without using GRAPE,
it offers a good speedup both for $O(N^2)$ direct summations and 
$O(N\log N)$ fast algorithms. The speedup factor for the latter is
about 10-100 times, depending on accuracy and other factors.

It seems obvious that the combination of the fast algorithm, parallel
computer and GRAPE hardware would offer a very high performance.
Kawai and Makino (2003, also see Makino 2004) reported implementations
of tree algorithms on PC-GRAPE cluster, which is a PC cluster with
each PC connected to a GRAPE hardware. As expected, such parallel
PC-GRAPE system achieved very high performance. For example, 
the largest simulation for a single virialized
$N$-body system (31 million particles) in the literature was performed
using 8-node GRAPE-5 cluster and parallel tree algorithm (Fukushige,
Kawai, Makino 2004).

In principle, by constructing a larger PC-GRAPE cluster we can
increase the size of the system we can handle.  However, current
hardwares, namely GRAPE-5 (Kawai et al. 2000) or GRAPE-6 (Makino et
al. 2003), are not suited for the construction of very large
GRAPE-Clusters. The reason is that they are too fast and also too
expensive. The reproduction cost of the smallest unit of GRAPE-6 is
significantly higher than the cost of one node of a PC cluster. On the
other hand, as discussed in Makino et al. (2003), the speed of the
host computer limits the speed of the tree algorithm. Thus, if we can
increase the number of host computers while keeping the money we spend
for GRAPE, we can achieve better performance for the tree algorithm.
This means we should develop a less expensive, small GRAPE hardware
compared to what is currently available.

We developed GRAPE-6A to achieve this goal. It utilizes the same
custom processor chip as was used for GRAPE-6. However, a single
GRAPE-6A card houses only four GRAPE-6 chips, compared to up to 32 chips
of single GRAPE-6 processor board. By limiting the number of chips to
four, we were able to design GRAPE-6A as a single PCI card, which can
fit directly into the host PCI bus. Thus, we successfully reduce the
cost of the smallest GRAPE-6 system by nearly a factor of ten.

The plan of this paper is as follows.  In sections 2 and 3, we
describe the GRAPE-6A hardware and software. In section 4, we discuss
the performance of a single GRAPE-6A hardware for three algorithms:
tree, TreePM, and individual timestep. In section 5 we describe a
medium-scale (24-node) PC-GRAPE cluster which we constructed. In
section 6, we discuss the performance of parallel algorithms on this
PC-GRAPE cluster. Section 7 is for discussion.

%%%%%%%%%%%%%%%%%%%%%%%%%%%%%%%%%%%%%%%%%%%%%%%
\section{GRAPE-6A Hardware}

\subsection{Function}
\label{sec:func}

GRAPE-6A calculates the gravitational force, its time derivative and
potential, given by
\begin{eqnarray}
{\vec a}_i & = &
\sum_j {m_j {\vec r}_{ij} \over (r_{ij}^2 + \varepsilon^2)^{3/2}} \\
{d{\vec a}_i \over dt} & = &
\sum_j m_j \left[{{\vec v}_{ij} \over (r_{ij}^2 + \varepsilon^2)^{3/2}}
+ {3({\vec v}_{ij}\cdot{\vec r}_{ij}){\vec r}_{ij} \over (r_{ij}^2
+ \varepsilon^2)^{5/2}} \right] \\
 \phi_i & = &
\sum_j {m_j \over (r_{ij}^2 + \varepsilon^2)^{1/2}}
\end{eqnarray}
where ${\vec a}_i$ and $\phi_i$ are the gravitational acceleration and
the potential of particle $i$, ${\vec r}_i$, ${\vec v}_i$, and $m_i$
are the position, velocity and mass of particle $i$, ${\vec r}_{ij} =
{\vec r}_{j} - {\vec r}_{i}$ and ${\vec v}_{ij} = {\vec v}_{j} - {\vec
v}_{i}$.  It also calculates predictor polynomials of position and
velocity for the individual timestep algorithm, evaluates the nearest
neighbor particle and distance, and constructs the list of neighbor
particles.

\subsection{Overall Structure}

\begin{figure}
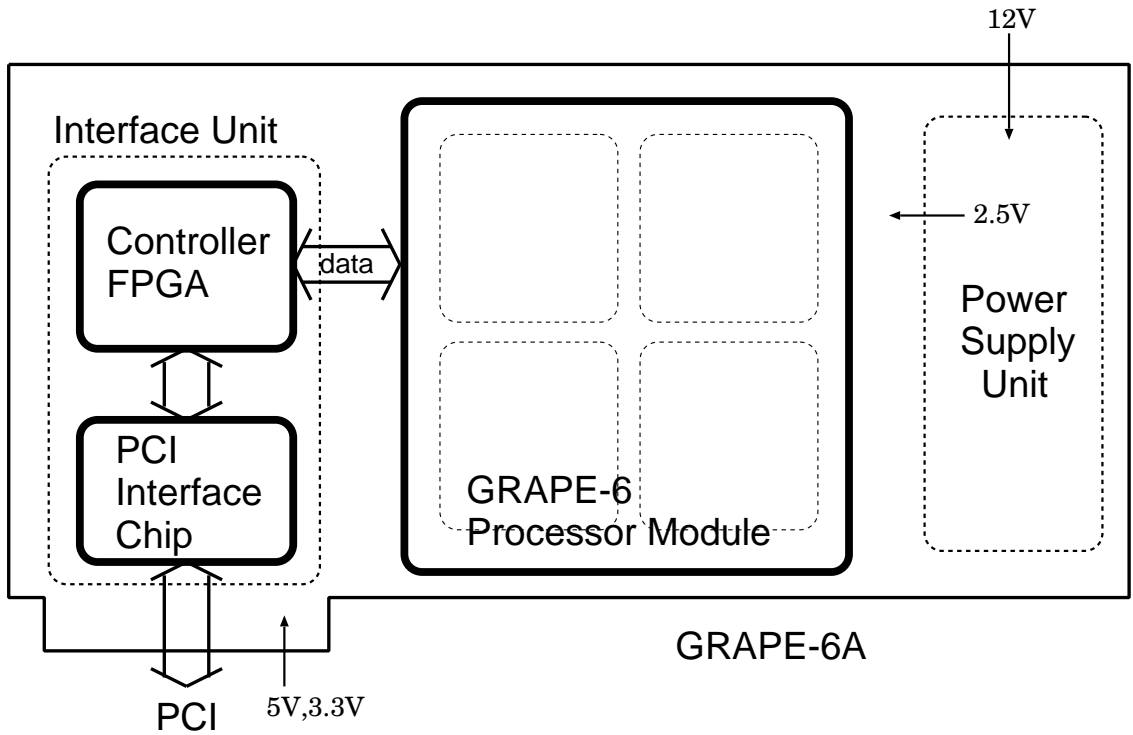

\begin{center}
\leavevmode
\FigureFile(15 cm,12 cm){./g6a.eps}
\end{center}
\caption{Overall structure of GRAPE-6A}
\label{g6a:fig1}
\end{figure}

Figure \ref{g6a:fig1} shows the overall structure of GRAPE-6A.  It is
a standard PCI short card onto which one GRAPE-6 processor module, an
interface unit, and a power supply unit are integrated. The processor
module comprises the custom processor chips for the functions
described in section \ref{sec:func}. The interface unit handles data transfer
between the host PCI bus and GRAPE-6 processor module. The power
supply unit converts the input voltage level to that of the GRAPE-6
processor module. We describe each unit in the following subsections.

\subsection{GRAPE-6 Processor Module}

\begin{figure}
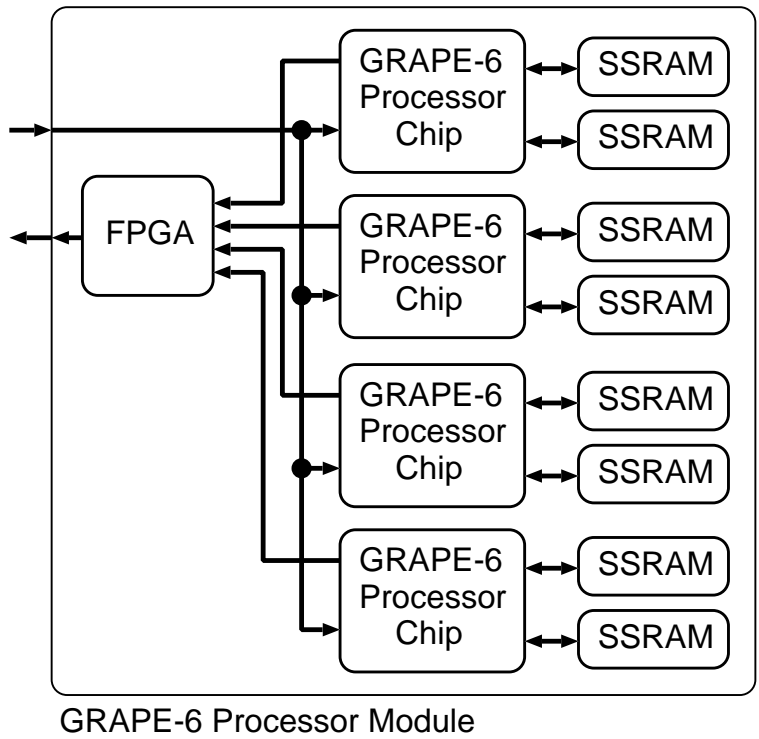

\begin{center}
\leavevmode
\FigureFile(10 cm,10 cm){./g6mod.eps}
\end{center}
\caption{Structure of GRAPE-6 processor module}
\label{g6mod:fig}
\end{figure}

A GRAPE-6 processor module consists of four GRAPE-6 processor chips,
eight SSRAM chips and one FPGA chip.  Figure \ref{g6mod:fig} shows its
structure.  A GRAPE-6 processor chip is a custom LSI chip dedicated to
the functions in section \ref{sec:func}. It consists of six force
calculation pipelines, a predictor pipeline, a memory interface, a
control unit and I/O ports. More details about GRAPE-6 processor chip
are described in Makino et al. (2003).  Two SSRAM chips are attached
to each GRAPE-6 processor chip. They are used to store the particle
data. The FPGA chip realizes a 4-input, 1-output reduction network for
the data transfer from GRAPE-6 processor chip to the host
computer.

Four GRAPE-6 processor chips calculate the forces on the same set of
($i$-)particles, but from a different set of ($j$-)particles (what is
called $j$-parallelization). The maximum number of 
$i$-particles is 48, since each chip has six real force calculation
pipelines and each real pipeline serves as eight virtual multiple
pipelines (VMP, Makino et al. 1997).  With VMP, one pipeline acts as
if it is  multiple
pipelines operating at a slower speed, each of which calculates the
force on its own $i$ particles, but from the same $j$ particles. Thus,
we can reduce the required memory bandwidth for $j$ particles.

The forces calculated by the four GRAPE-6 processor chips are summed
up in the FPGA chip on GRAPE-6 processor module and sent back to the
host computer.  Maximum numbers of $j$-particles that can be stored in
the memory is 16384 per chip, and 65536 for the GRAPE-6 processor
module we used. A newer version of the processor module which can keep
up to 131072 particles is commercially available.

The peak speed of one GRAPE-6 processor module, $i.e.$, GRAPE-6A, is
131.3 Gflops (force and its derivative)/ 87.5 Gflops (force only).  It
has 24 (real) pipelines and operates at 96 MHz.  Here we count
operations for gravitational force and its time derivatives and those
for force only as 57 and 38 floating point operations, respectively.
The clock signal for GRAPE-6 processor module is generated and
supplied by a crystal oscillator on board. Input clock frequency for
the processor chip is 24 MHz. It generates the clock signal with four
times higher frequency in on-chip PLL (Phase Lock Loop) circuit, and
all logic in the processor chips and SSRAM chips operate on this
multiplied clock.

\subsection{Interface Unit}

The interface unit consists of a PCI interface chip and a controller
FPGA chip. For the PCI interface chip, we used the PCI 9080 chip from
PLX technology Inc., which we used also for GRAPE-4A, 5 and 6. The
interface to the host computer is standard 32bit/33MHz PCI bus. The
controller FPGA chip handles data transfer between the PCI interface
chip and GRAPE-6 processor module.  For the controller FPGA chip, we
used Altera EP1K100FC256.  The function and performance of the
interface unit are basically the same as those of the PCI Host
Interface Board (Kawai et al. 1997), which is used for GRAPE-4 and
GRAPE-5.

\subsection{Power Supply Unit}

GRAPE-6A also has the power supply unit on board.  The power supply
unit converts the 12V power supplied from the power supply unit of the
host PC to 2.5V required by the processor chip and other chips.
The design of the power supply unit is essentially the same as that of
(relatively old) PC motherboard for x86 processors. In fact, the
circuit design is that same as that of the reference design of the
controller chip we used (LTC1709-8 from Linear Technology).

The 12V power supply comes from an additional power connector, which
accepts a standard 4-pin connector for  hard-disk units. In
addition to the 2.5V supply, GRAPE-6A requires 3.3V and 5V supplies as
well. For these, the supply through the PCI connector is sufficient.

\subsection{Physical Design}

\begin{figure}
\begin{center}
\leavevmode
\FigureFile(10 cm,8 cm){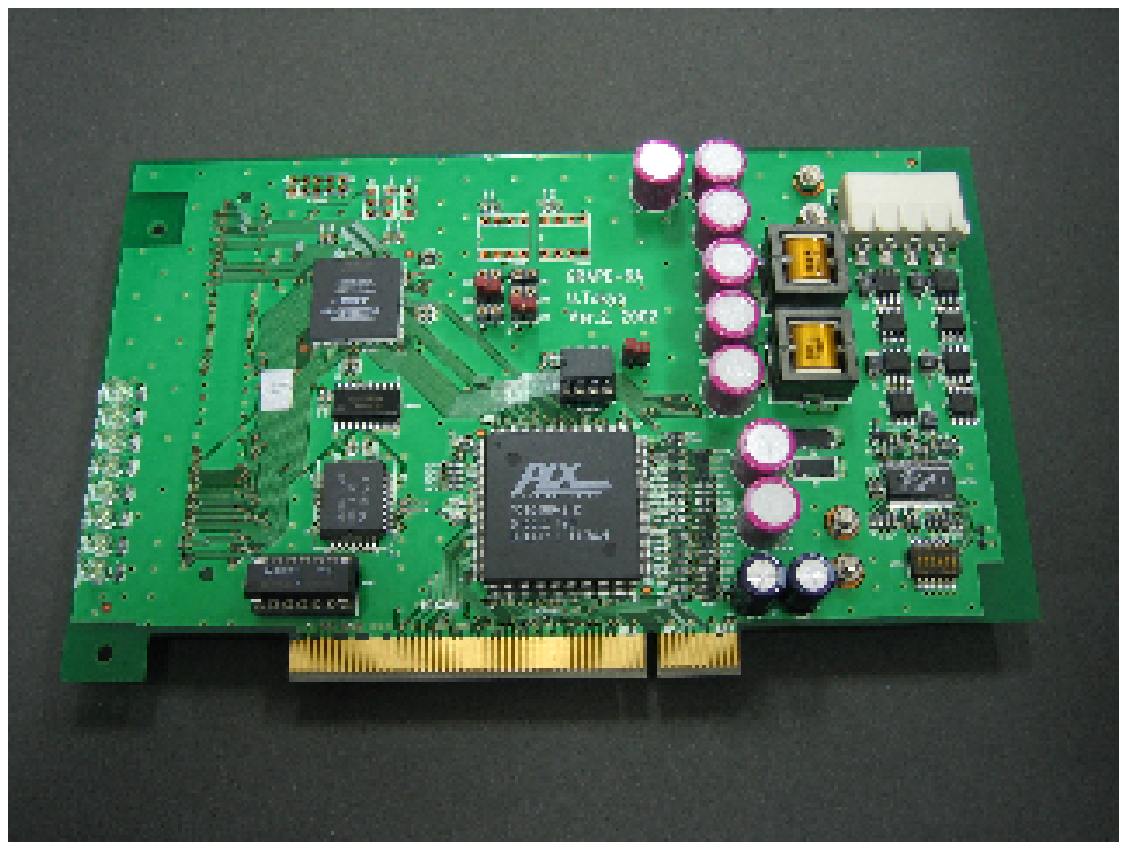}
\FigureFile(10 cm,8 cm){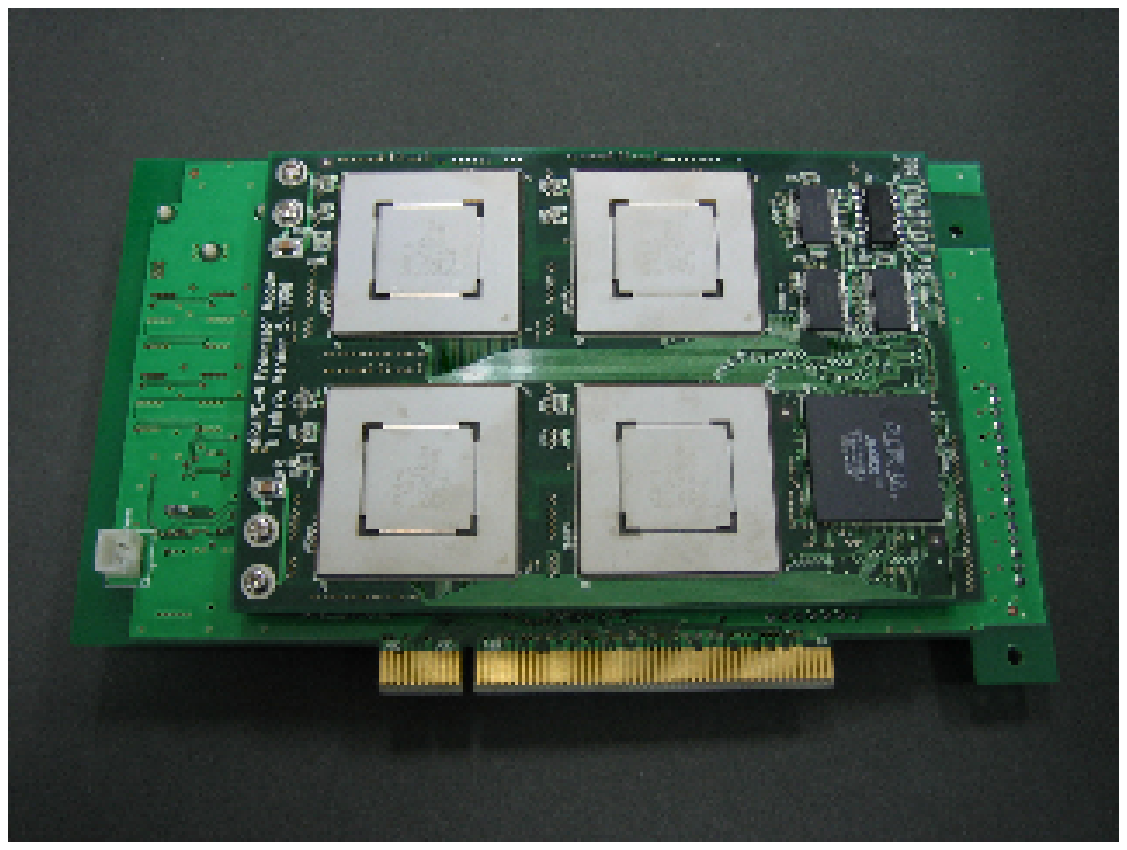}
\end{center}
\caption{Top and bottom views of a GRAPE-6A board}
\label{g6a:fig2}
\end{figure}

Figure \ref{g6a:fig2} shows the top (upper panel) and bottom views
(lower panel) of a GRAPE-6A board.  The interface unit and the power
supply units are on the left and right sides of the top view,
respectively. The GRAPE-6 processor module mounted on the bottom side.
The board is 8 layer PCB (Printed Circuit Board), and its size is
roughly 11 cm by 17 cm (standard short PCI card).  The board design
began on the autumn of 2001, and it completed on January, 2003 after
once redesign of PCB.  A commercial version of GRAPE-6A (called
MicroGRAPE) is now available.

\subsection{Difference from GRAPE-6}

The primary difference between GRAPE-6A and GRAPE-6 is in the number
of GRAPE-6 processor chips. GRAPE-6A has 4 chips, while a single-board
GRAPE-6 can house up to 32 chips, and multi-board configuration
connected to a single host is available.  As we stated in the
introduction, for GRAPE-6A we intentionally reduced the maximum number
of chips, so that the total system of multiple host computers and
multiple GRAPE-6A system offer the performance higher than that of a
single GRAPE-6 system with the same number of chips, at least for
approximate algorithms such as the tree algorithm.  We discuss the
performance difference between GRAPE-6 and GRAPE-6A in sections
\ref{perf:tree} and \ref{sec:its}.

\section{Interface Software}

The design principle of the interface software of GRAPE-6A is the same
as that of previous GRAPE systems (Makino, Funato 1993, Makino, Taiji
1998 subsection 5.3, Kawai et al. 2000). Low-level software that
communicates with GRAPE-6A is encapsulated into the user library
functions and is hidden to the user. The user program accesses
GRAPE-6A only through the library functions. The API(Application
Program Interface) of the GRAPE-6A library is designed so as to be
same as that of the GRAPE-6 software library. Another software library
whose API is the same as that of the GRAPE-5 software library is
prepared. The GRAPE-6A software library is available on a web
site\footnote{{\tt http://grape.c.u-tokyo.ac.jp/}\~{\tt fukushig/g6a}}.

\section{Performance of single GRAPE-6A for practical algorithms}

In this section, we describe the implementation and measured
performance of three force calculation algorithms, Barnes-Hut tree
algorithm, TreePM algorithm and individual timestep algorithm, on a
single board of GRAPE-6A. Except for the performance measurement in
Table \ref{per:tab1} and \ref{its:tab1}, we used a host computer with
an Intel Pentium 4 processor (2.8CGHz, i865G) and 2GB of PC3200
memories, and {\tt gcc} complier (version 3.2.2) on RedHat 9.0 (Linux
kernel 2.4.20-8).

\subsection{Tree algorithm}

The Barnes-Hut tree algorithm (Barnes, Hut 1986) reduces the
calculation cost from $O(N^2)$ to $O(N\log N)$, by replacing the
forces from distant particles by that from a virtual particle at their
center of mass (or multiple expansion). In order to use efficiently
the GRAPE hardware, we use the modified tree algorithm developed by
Barnes (1990) and first used on GRAPE-1A by Makino (1991a).  With this
algorithm the tree traversal is performed for a group of neighboring
particles and an interaction list is created.  GRAPE calculates the
force from particles and nodes in this interaction list to particle in
the group.  In the original algorithm, tree traversal is performed for
each particle. Since GRAPE cannot perform the tree traversal
algorithm, the tree traversal must be done on the host computer. By
reducing the number of tree traversal operations, the modified
algorithm greatly reduce the work of the host computer and improves
the overall performance. This modified algorithm was originally
developed for a vector processor (CDC Cyber 205) and has been used on
general-purpose MPP (Dubinski 1996).

\subsubsection{Calculation procedure}

With the modified tree algorithm, the GRAPE-6A system performs the
integration of one time step in the following way. This procedure is
the same as that for  other GRAPEs:
\begin{itemize}
\item[1.] The host computer constructs a tree structure. 

\item[2.] Repeat steps 3 through 7 until all the forces on all
particles are updated.

\item[3.] Identify a group of particles for which the same interaction
list is used from remaining particles. This part is done by traversing
the tree structure.

\item[4.] The host computer creates the interaction list for a group, 
and sends the data of the particles listed up to GRAPE-6A. 

\item[5.] Repeat steps  6 and 7 for all particles in a group. 

\item[6.] The host computer sends particles to be calculated to GRAPE-6A. 

\item[7.] GRAPE-6A calculates the forces exerted on the particles, and 
then returns the result to the host computer. 

\item[8.]  The host computer updates the positions and velocities of all
particles using the calculated force. 

\end{itemize}

The modified tree algorithm reduces the calculation cost of the host
computer by roughly a factor of $n_{\rm g}$, where $n_{\rm g}$ is the
average number of particles in groups.  On the other hand, the amount
of work on GRAPE increases as we increase $n_{\rm g}$, since the
interaction list becomes longer.  There is, therefore, an optimal
value of $n_{\rm g}$ at which the total computing time is the minimum
(Makino 1991a).  The optimal value of $n_{\rm g}$ depends on various
factors, such as the relative speed of GRAPE and its host computer,
the opening parameter and the number of particles.  For the present
GRAPE-6A, $n_{\rm g}=500-1000$ is close to optimal.

\subsubsection{Performance}
\label{perf:tree}

\begin{figure}
\begin{center}
\leavevmode
\FigureFile(10 cm,8 cm){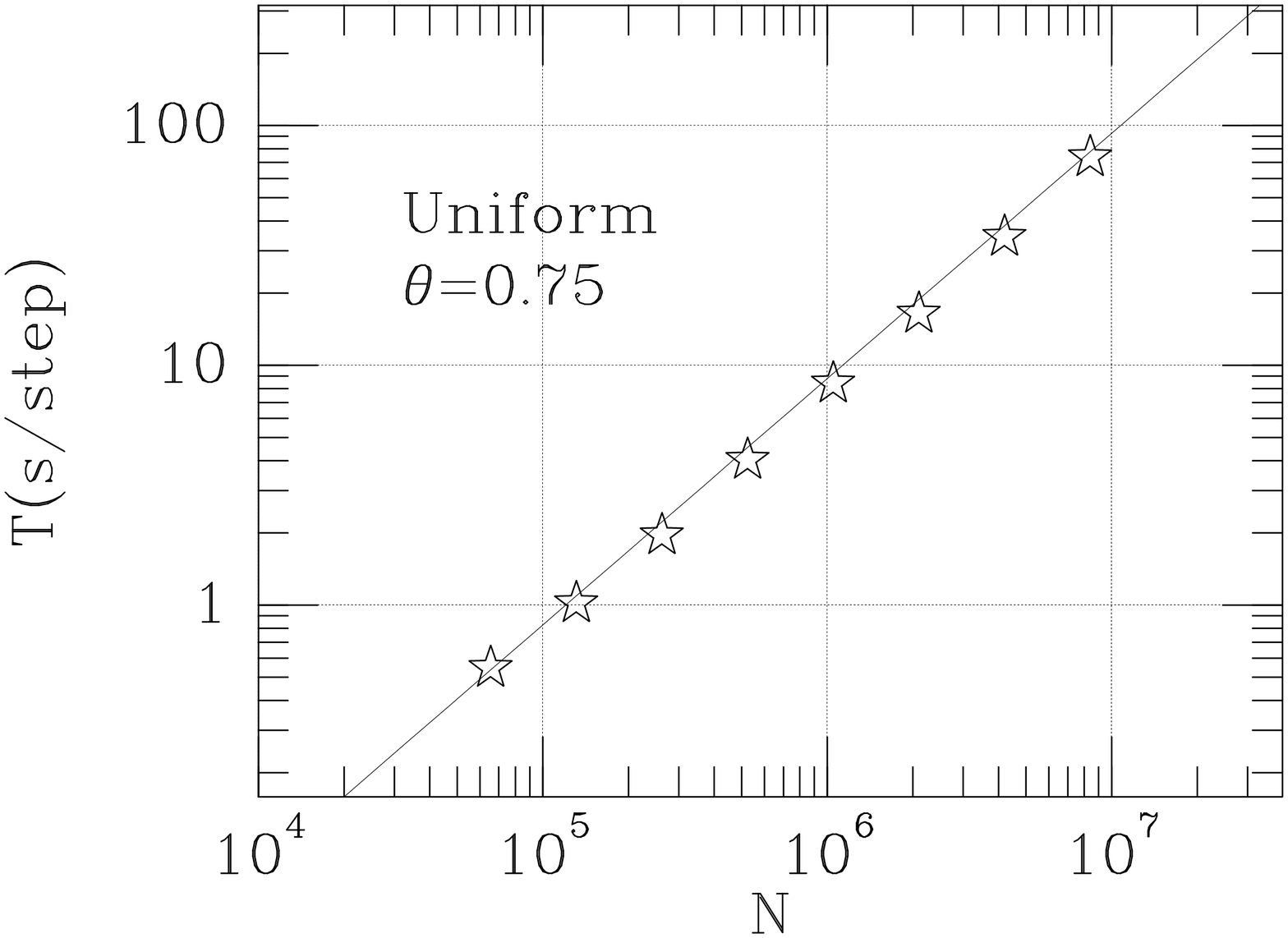}
\end{center}
\caption{Calculation time for the Barnes-Hut tree algorithm
($\theta=0.75$) as a function of number of particles $N$ . The solid line
indicates the time estimated with the theoretical model given in the text.}
\label{per:fig1}
\end{figure}

\begin{figure}
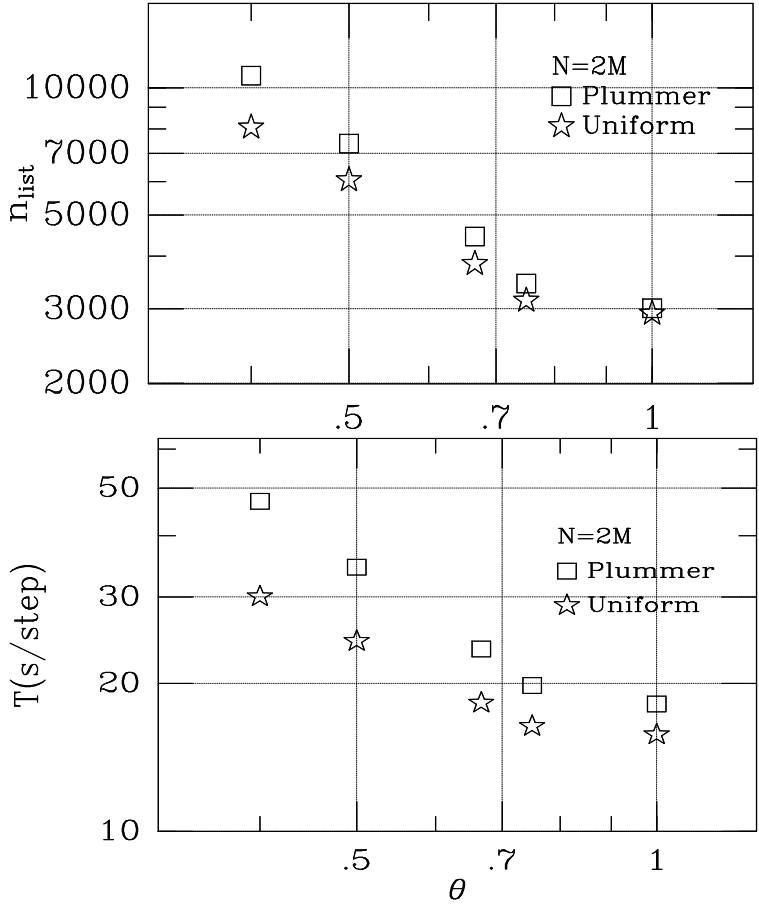

\begin{center}
\leavevmode
\FigureFile(10 cm,12 cm){./treemon23.eps}
\end{center}
\caption{Average length of the interaction list and calculation
time for Barnes-Hut tree algorithm as a function of the opening angle
$\theta$ for the uniform sphere (top panel) and the Plummer model
(bottom panel). The number of
particles is 2097152. }
\label{per:fig2}
\end{figure}

Figure \ref{per:fig1} shows the measured
calculation time per one timestep as a function of the number of
particle, $N$.  For the particle distribution, we use an uniform sphere
of equal mass particles.  The opening angle, $\theta=0.75$ (dipole
expansion) and $n_g \simeq 500-1000$. We set
$n_{\rm crit}=2000$, where $n_{\rm crit}$ is the maximum number of
particles in the group.  We can see that the calculation time grows
practically linearly as we increase $N$.

Figure \ref{per:fig2} shows the measured calculation time and average
length of the interaction list as a function of the opening angle
$\theta$ for the uniform sphere and the Plummer model (the cutoff
radius is 22.8 in a system of units such that $M=G=-4E=1$, where $E$
is the total energy).  The number of particles is $N=2097152$.  We can
see the dependence of $\theta$ is weak for $\theta \ge 0.7$.  This is
because the length of the interaction list does not change much for
$\theta \ge 0.7$, which is a characteristic of GRAPE implementation of
the modified tree algorithm (Makino 1991a).  We can also see that the
calculation time and the average length of the interaction list for
Plummer model are longer than those for the uniform sphere. 

In the following, we present a theoretical model for the performance.
The total calculation time per timestep is expressed as
\begin{equation}
T = T_{\rm host} + T_{\rm grape} + T_{\rm comm},
\end{equation}
where $T_{\rm host}$, $T_{\rm grape}$, and $T_{\rm comm}$ are the time
spent on the host computer, the time spent on GRAPE-6A, and the time
spent for data transfer between the host computer and GRAPE-6A,
respectively. The time spent on the host computer is expressed as
\begin{equation}
T_{\rm host} = (N\log_{10}N) t_{\rm const} + n_{\rm list}{N\over n_{\rm g}}t_{\rm list},
\end{equation}
where $t_{\rm const}$ and $t_{\rm list}$ are the times to construct
the tree structure and the interaction lists, respectively. In this
equation $n_{\rm list}$ is the average length of the interaction
list. According to Makino (1991a), $n_{\rm list}$ can be estimated as
follows:
\begin{equation}
n_{\rm list} \simeq n_{\rm g} + 14n_{\rm g}^{2/3}
+ 84 n_{\rm g}^{1/3} + 56\log_8 n_{\rm g} -31
\theta^{-3}\log_{10} n_{\rm g} -72 -100
\theta^{-3}\log_{10}{N\theta^3\over 23}.
\end{equation}
The time spent on GRAPE-6A is expressed as
\begin{equation}
T_{\rm grape} = N n_{\rm list} t_{\rm pipe},
\end{equation}
where $t_{\rm pipe}$ is the time to calculate one pairwise interaction
on GRAPE-6A. The time spent for data transfer is expressed as
\begin{equation}
T_{\rm comm} = 24n_{\rm list}{N\over n_{\rm g}}t_{{\rm comm},j}
+ 28Nt_{{\rm comm},i} + 56N t_{{\rm comm},f}.
\end{equation}
Three terms in the right-hand side indicate the times for data
transfer of $j$-particles, $i$-particles, and calculated forces. Here,
$t_{{\rm comm},j}$, $t_{{\rm comm},i}$, and $t_{{\rm comm},f}$ are the
times to transfer one byte data between the host computer and
GRAPE-6A, for the $j$-particle, the $i$-particle, and the calculated
force, respectively. These times include not only time for transfer
through PCI but also that for conversion of data from/to conventional
floating format in the host computer to/from number formats in GRAPE-6
chip.  The calculation time estimated using the theoretical model is
plotted in Figure \ref{per:fig1}. Here, we set $n_{\rm g}=800$ and the
time constants as $t_{\rm const}=1.4\times 10^{-7}$ (s), $t_{\rm
list}=2.3\times 10^{-7}$ (s), $t_{\rm pipe}=4.3\times 10^{-10}$ (s),
and $t_{{\rm comm},j}=t_{{\rm comm},i}=t_{{\rm comm},f}=2.7\times
10^{-8}$ (s).

\begin{table}
\caption{Calculation times per timestep of the tree algorithm on
various host computers  ($N=2097152, \theta=0.75$)} 
\begin{center}
\begin{tabular}{llc|ccc|l}
\hline
\hline
CPU        & chipset  & memory & $T$(s) & $T_{\rm host}$(s) & $T_{\rm comm}$(s) & note\\
\hline
Athlon64 3500+ & K8T800 & 6.4GB/s & 14.2 & 2.6 & 8.6 &\\
Pentium4 2.8CGHz & i865G & 6.4GB/s & 15.6 & 3.7 & 9.0 & PCI 40MHz\\
Pentium4 3.0EGHz & i865G & 6.4GB/s & 15.9 & 3.3 & 9.4 & \\
Pentium4 2.8CGHz & i865G & 6.4GB/s & 16.4 & 3.7 & 9.8 & \\
Xeon 3.0DGHz & E7525 & 5.4GB/s & 16.4 & 3.6 & 9.7 & \\
Opteron 244 & 8111 & 5.4GB/s & 18.0 & 2.6 & 12.2 &\\
Pentium4 2.4BGHz & E7205 & 4.2GB/s & 18.1 & 4.6 & 10.3 &\\
Pentium4 550 & i915G & 6.4GB/s & 18.5 & 3.1 & 12.3 & \\
\hline
\end{tabular}
\end{center}
\label{per:tab1}
\end{table}

Table \ref{per:tab1} shows the performance of the tree algorithm with
GRAPE-6A for several different host computers. 
The calculation time per
timestep for an uniform sphere of $N=2097152$ is shown. We set $\theta=0.75$
and $n_{\rm crit}=2000$. For the results shown in tables \ref{per:tab1}
and \ref{its:tab1}, we used the {\tt gcc} complier, version 3.2.2 on RedHat
9.0(Linux kernel 2.4.20-8) for the host computers with Pentium 4, and
{\tt gcc} complier, version 3.3.3 on Fedora Core 2.0 (Linux kernel
2.6.5-1) for other host computers.  Table \ref{per:tab1} shows that
total calculation time is determined mainly by the time for
communication. In particular, Opteron+8111 and Pentium550+i915G shows
rather low communication speed. This result is consistent with the
relatively low PCI DMA performance of these machines. 

\begin{table}
\caption{Calculation time per timestep for the tree algorithm on
GRAPE-6A, -5 and -6 ($N=2097152, \theta=0.75$).  Peak speeds in
(pairwise) interactions per second are also listed.}
\begin{center}
\begin{tabular}{lc|ccc|cccc}
\hline
\hline
       & Peak(int/s) & $n_{\rm crit}$ & $n_{\rm g}$ & $n_{\rm list}$
       & $T$(s) & $T_{\rm host}$(s) & $T_{\rm grape}$(s) & $T_{\rm comm}$(s) \\
\hline
GRAPE-6A         & $2.30 \times 10^9$ & 2000 & 836 & 3145
& 16.4 & 3.7 & 2.9 & 9.8 \\
GRAPE-5          & $1.28 \times 10^9$ & 2000 & 836 & 3145
& 16.4 & 3.6 & 5.3 & 7.3 \\
GRAPE-6(8 chips) & $4.41 \times 10^9$ & 6000 & 1125 & 4393
& 20.2 & 3.6 & 2.1 & 14.3 \\
GRAPE-6(16 chips) & $8.83 \times 10^9$ & 8000 & 5053 & 12671
& 17.1 & 2.9 & 3.0 & 11.0 \\
GRAPE-6(24 chips) & $13.24 \times 10^9$ & 8000 & 5053 & 12671
& 16.1 & 2.9 & 2.0 & 11.0 \\
\hline
\end{tabular}
\end{center}
\label{per:tab2}
\end{table}

Table \ref{per:tab2} shows the calculation times per timestep for an
uniform sphere of $N=2097152$ ($\theta=0.75$, $n_{\rm crit}=2000$) for
GRAPE-6A, 5 and 6.  The relative speed of these hardware is
practically independent of $N$, since $T_{\rm host}$, $T_{\rm grape}$
and $T_{\rm comm}$ all depend almost linearly on $N$. From table
\ref{per:tab2} we can see that the performance of GRAPE-6A is almost
the same as that of GRAPE-5, and is not slower than that of
GRAPE-6. In other words, GRAPE-6A, with one GRAPE-6 module, has
sufficient performance for the tree algorithm.  Note that the
difference in the communication performance among GRAPE-6A, GRAPE-5,
and GRAPE-6 comes from the amount of data to be transferred. The
amount of data for one particle is the smallest for GRAPE-5, since it
uses the word formats shorter than those used in GRAPE-6. Though
GRAPE-6A and GRAPE-6 use the same data format, amount of data transfer
is smaller for GRAPE-6A since we developed a highly optimized library
functions which minimizes the data transfer for GRAPE-6A.  For
GRAPE-6, we used the library function designed to be used with
individual timestep algorithm with Hermite integration scheme.

\subsection{TreePM algorithm}
\label{sec:treepm}

In this section, we briefly describe an implementation of algorithm
for simulations with periodic boundary condition on GRAPE-6A. Periodic
boundary has been widely used for studies of cosmological structure
formation. We implemented the TreePM algorithm (Bagla and Ray 2003) on
GRAPE-6A. In the TreePM algorithm, the force is split into two
components: a long-range force calculated using particle-mesh (PM)
technique and a short-range force calculated using the tree
algorithm. In other words, TreePM is a variant of the $\rm P^3M$
algorithm, where we use a tree algorithm for the evaluation of the
particle-particle, short-range force. It solves the problem of $\rm
P^3M$ scheme that the calculation cost goes up quickly as the system
become inhomogeneous.

The tree part uses GRAPE-6A in the way same as that for open boundary
problem. GRAPE-6A can calculate the $r^{-2}$ force multiplied by a
user-specified cutoff function.

\begin{table}
\caption{Calculation time per timestep for the TreePM algorithm}
\begin{center}
\begin{tabular}{cc|ccc}
\hline
\hline
$N$ & $z$ & $T$(s) & $T_{\rm PM}$(s) & $T_{\rm tree}$(s) \\
\hline
$64^3$ & 24 & 2.1 & 0.5 & 1.6 \\
$128^3$ & 28 & 17.3 & 3.9 & 13.3 \\
$128^3$ & 0 & 16.9 & 3.8 & 13.1 \\
\hline
\end{tabular}
\end{center}
\label{per:tab3}
\end{table}

Table \ref{per:tab3} summarize the calculation time per timestep.  
Here, $z$ is the redshift, and the times, $T$, $T_{\rm PM}$, and
$T_{\rm tree}$ are the total time, the time spent for the PM force,
and the time spent for the tree force, respectively.  We used particle
distributions from dark matter simulations in LCDM cosmology whose box
size is 75Mpc. The opening angle $\theta=1.0$ and $n_{\rm crit}=2000$,
and the grid size for the PM force is set to $L/N^{1/3}$, where $L$ is
the box size.  A remarkable feature of the TreePM algorithm on GRAPE
is that the calculation time is almost constant during the
simulation. For more details, see Yoshikawa and Fukushige (2005).

\subsection{Individual timestep algorithm}

GRAPE-6A can also accelerate the individual timestep algorithm, which
is the original purpose of the GRAPE-6 development.  The individual
timestep algorithm is an extremely powerful tool for studies of dense
stellar systems. The basic idea of the individual timestep algorithm
is to assign different times and timesteps to particles in the
system. In order to use efficiently the GRAPE hardware, we use the
block individual timestep algorithm (McMillan 1986, Makino 1991b), in
which the system time is quantized to 2's powers so that multiple
particles have updated exactly at the same time,  and the fourth-order
Hermite integration scheme (Makino 1991c, Makino, Aarseth 1992). For more
details, see Makino et al.  (1997) or Makino and Taiji (1998).

\subsubsection{Calculation procedure}

With the individual timestep algorithm, the GRAPE-6A system performs the
integration of one time step in the following way: 

\begin{itemize}
\item[1.] As the initialization procedure, the host computer
sends all data of all particles to the memory on GRAPE-6A (in
the processor module). 

\item[2.] The host computer creates the list of particles to be
integrated at the present timestep. 

\item[3.] Repeat 4.-6. for all particles in the list. 

\item[4.] The host computer predicts the position and velocity of the
particle, and sends them to GRAPE-6A. 

\item[5.] GRAPE-6A calculates the force from all other 
particles, and then returns the results to the host computer. 

\item[6.] The host computer integrates the orbits 
of the particles and determines new timestep. The updated
particle data are sent to the memory on GRAPE-6A. 

\item[7.]  The host computer updates 
the present system time and go back to step 2. 

\end{itemize}

With the individual timestep algorithm, the maximum number of
particles one can use is limited by the amount of memory available on
the side of the GRAPE hardware.  With the present GRAPE-6 processor
module, the limit is 65536 (resent commercial-version hardwares can
store up to 131072). This number can be a bit small for some
applications. 

The simplest way to overcome this limitation is 
to use multiple GRAPE-6A boards for single calculation. We can achieve
this goal by either putting multiple GRAPE-6A boards to a single host
or by using a cluster of host+GRAPE-6A systems. For the tree
algorithm, a cluster of multiple host computers each with single
GRAPE-6A card  is preferred over a single host with multiple
GRAPE-6A. So we will discuss the performance of parallel host system
in section \ref{sec:pits}.

Note that only the individual timestep algorithm with
direct-summation  suffers this limitation of the memory size.
With other
algorithm, we can handle the number of particles larger than the limit
by dividing the particles into subgroup and summing up the partial
forces from each subgroup. Also, in the case of tree-based algorithms,
the length of the interaction list is usually much smaller than
65,536 for any number of particles.

\subsubsection{Performance}
\label{sec:its}

As the benchmark runs, we integrated the
Plummer model with equal-mass particles for one time unit.  We use the
standard unit (Heggie and Mathieu 1986) in which  $M=G=-4E=1$. Here
$M$ and $E$ are the total mass and energy of the system, and $G$ is
the gravitational constant. The timestep criterion is that of Aarseth
(1999) with $\eta=0.01$. For the softening parameter, we used a
constant softening, $\varepsilon=1/64$ and a $N$-dependent softening,
$\varepsilon=4/N$. We performed the simulation for time 0 to time 2,
and used the calculation time for time 1 to time 2 to avoid the
complication due to the startup procedure.

\begin{figure}
\begin{center}
\leavevmode
\FigureFile(10 cm,8 cm){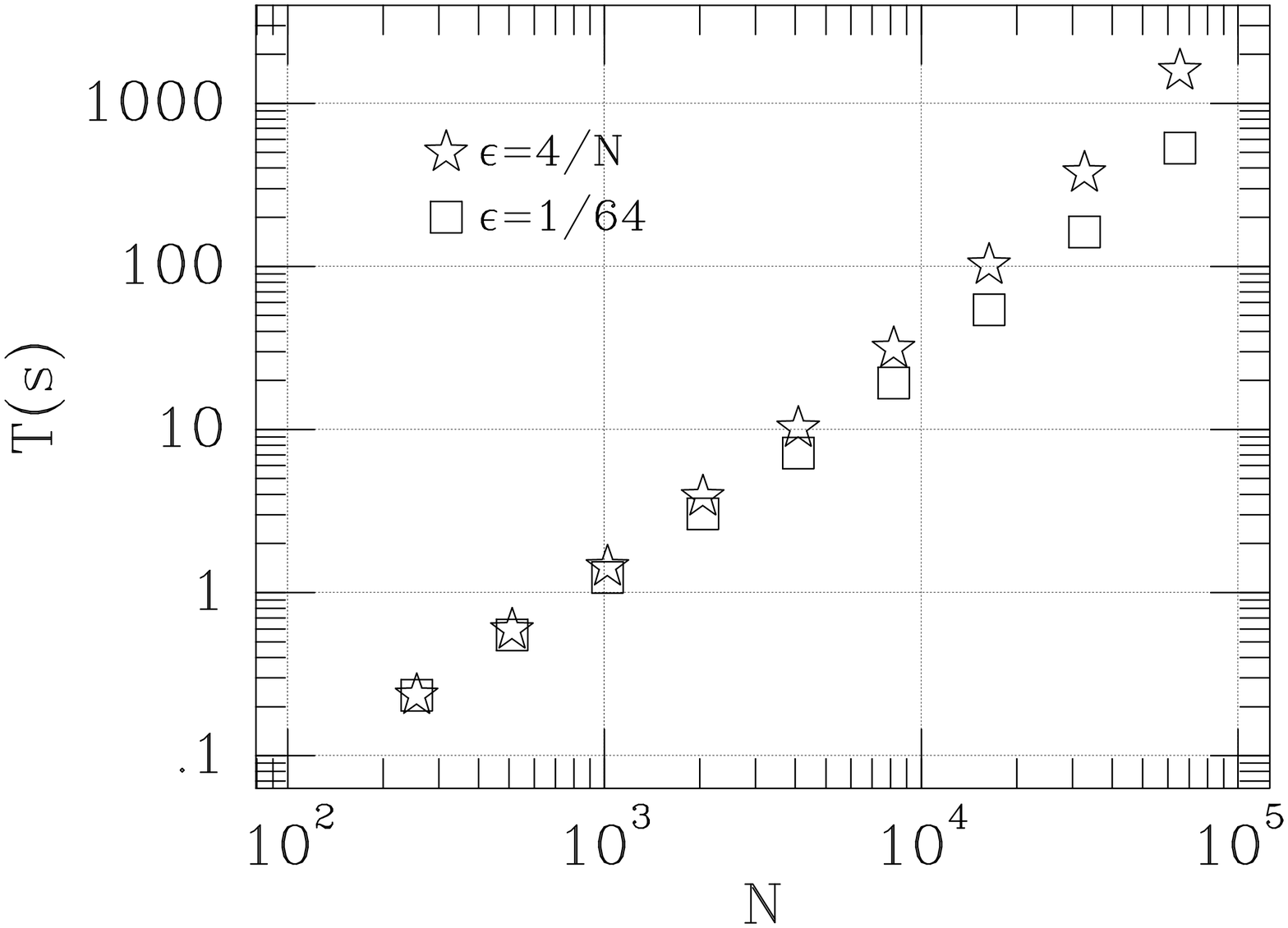}
\end{center}
\caption{Calculation time per unit time for the individual
timestep algorithm as a function of the number of particles $N$. Stars
and squares  indicate the result with softening length $\varepsilon = 4/N$
and $1/64$, respectively.}
\label{its:fig1}
\end{figure}

\begin{figure}
\begin{center}
\leavevmode
\FigureFile(10 cm,8 cm){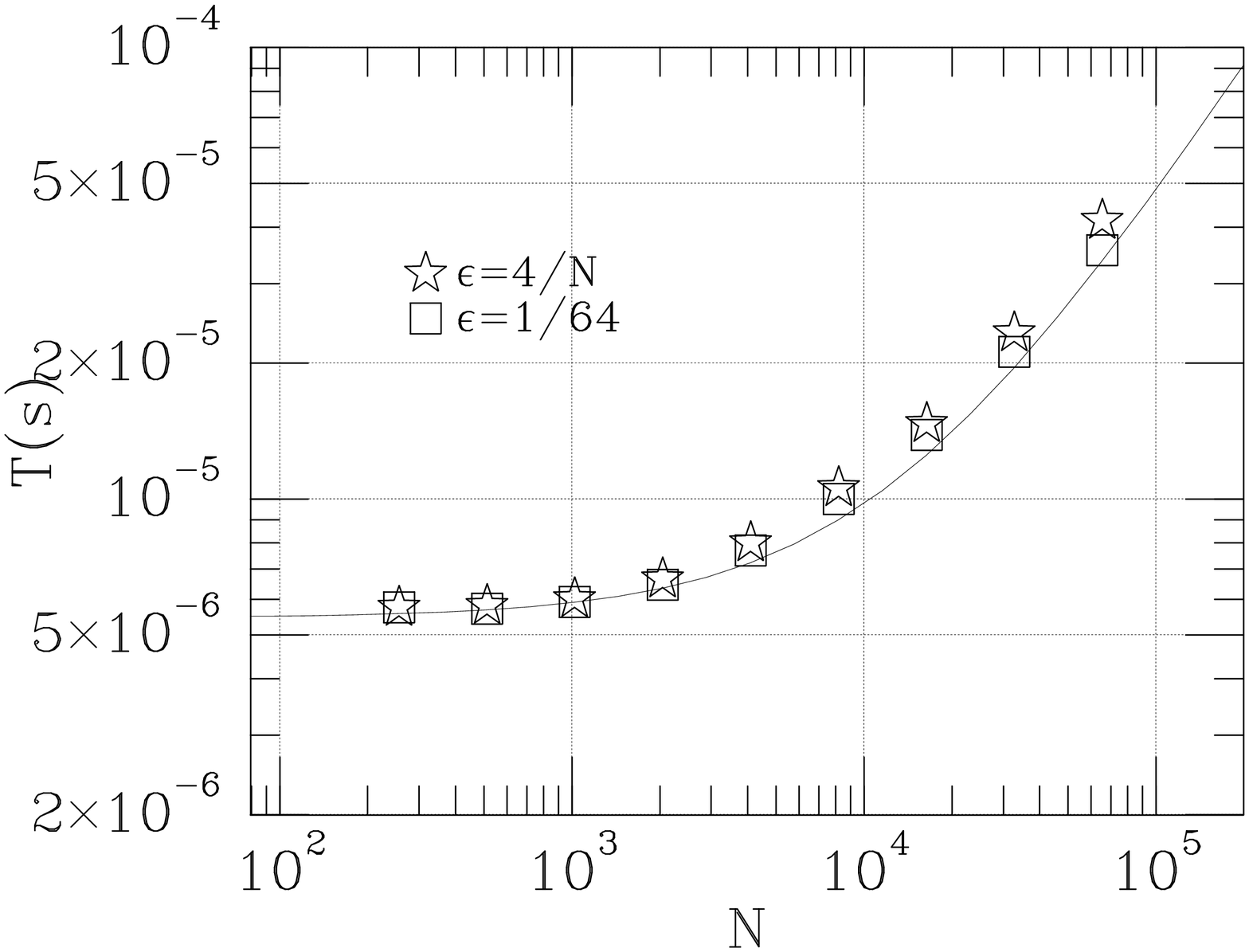}
\end{center}
\caption{Calculation time for one particle step of the individual
timestep algorithm as a function of number of particles $N$. The solid
curve indicates the time estimated with the theoretical model given in the
text.}
\label{its:fig2}
\end{figure}

Figure \ref{its:fig1} shows the calculation time to integrate the
system for one time unit as a function of the number of particle, $N$.
The calculation time for the runs with the $N$-dependent softening is
longer than that for the constant softening, because the number of
timesteps is larger.  Figure \ref{its:fig2} shows the calculation time
per one particle step as a function of $N$. We can see that the time
per step is almost independent of the choice of softening.

We give a theoretical model for the performance as follows.  The
calculation time per one particle step is expressed as
\begin{equation}
T = t_{\rm host} + N t_{\rm pipe} + t_{\rm comm}, 
\end{equation}
Here $t_{\rm host}$ is the time for host computer to perform
computations to integrate one particle. 
The second term of the right-hand side is the time to calculate the
force and its time derivative for one particle on GRAPE-6A.
The third term, $t_{\rm comm}$, expresses the time to transfer data,
given by  
\begin{equation}
t_{\rm comm} = 60 t_{{\rm comm},i} + 56 t_{{\rm comm},f} + 72 t_{{\rm.comm},j} 
\end{equation}
where the first, second, and third terms are the times to transfer
data at step 4. 5. and 6. in the procedure, respectively.  These times
include the data conversion.  The calculation time per one particle
step estimated using the theoretical modeling are plotted in Figure
\ref{its:fig2}.  Here, we set the time constants as $t_{\rm
host}=3.8\times 10^{-7}$ (s), $t_{\rm pipe}=4.3\times 10^{-10}$ (s),
and $t_{{\rm comm},j}=t_{{\rm comm},i}=t_{{\rm comm},f}=2.7\times
10^{-8}$ (s).

\begin{table}

\caption{Calculation speed and time constants of the individual
timestep algorithm for various host computers}
\begin{center}
\begin{tabular}{llc|ccc|l}
\hline
\hline
CPU        & chipset  & memory 
& $t$($N$=1k)(s) & $t_{\rm host}$(s) & $t_{\rm comm}$(s) & note\\
\hline
Athlon 64 3500+ & K8T800 & 6.4GB/s
& $5.00 \times 10^{-6}$ & $2.74 \times 10^{-7}$ & $4.22 \times 10^{-6}$ &\\
Xeon 3.0D & E7525 & 5.4GB/s
& $5.58 \times 10^{-6}$ & $3.32 \times 10^{-7}$ & $4.80 \times 10^{-6}$ & \\
Pentium 4 2.8C & i865G & 6.4GB/s 
& $5.99 \times 10^{-6}$ & $4.21 \times 10^{-7}$ & $5.08 \times 10^{-6}$ & PCI 40MHz\\
Pentium 4 2.8C & i865G & 6.4GB/s 
& $6.02 \times 10^{-6}$ & $4.24 \times 10^{-7}$ & $5.10 \times 10^{-6}$ & \\
Pentium 4 3.0E & i865G & 6.4GB/s 
& $6.45 \times 10^{-6}$ & $4.52 \times 10^{-7}$ & $5.51 \times 10^{-6}$ & \\
Pentium 4 2.4B & E7205 & 4.2GB/s 
& $6.84 \times 10^{-6}$ & $4.83 \times 10^{-7}$ & $5.84 \times 10^{-6}$ &\\
Opteron 244 & 8111 & 5.4GB/s 
& $7.35 \times 10^{-6}$ & $3.18 \times 10^{-7}$ & $6.53 \times 10^{-5}$ & \\
Pentium 4 550 & i915G & 6.4GB/s 
& $7.63 \times 10^{-6}$ & $3.52 \times 10^{-7}$ & $6.81 \times 10^{-6}$ & \\
\hline
\end{tabular}
\end{center}
\label{its:tab1}
\end{table}

Table \ref{its:tab1} shows the performance on various host computers
for $N=1024$. We used relatively small $N$ to see the different in the
speed of host computers. Here again, the speed of the host computer is
limited mainly by the speed of data conversion and communication.

\begin{figure}
\begin{center}
\leavevmode
\FigureFile(10 cm,8 cm){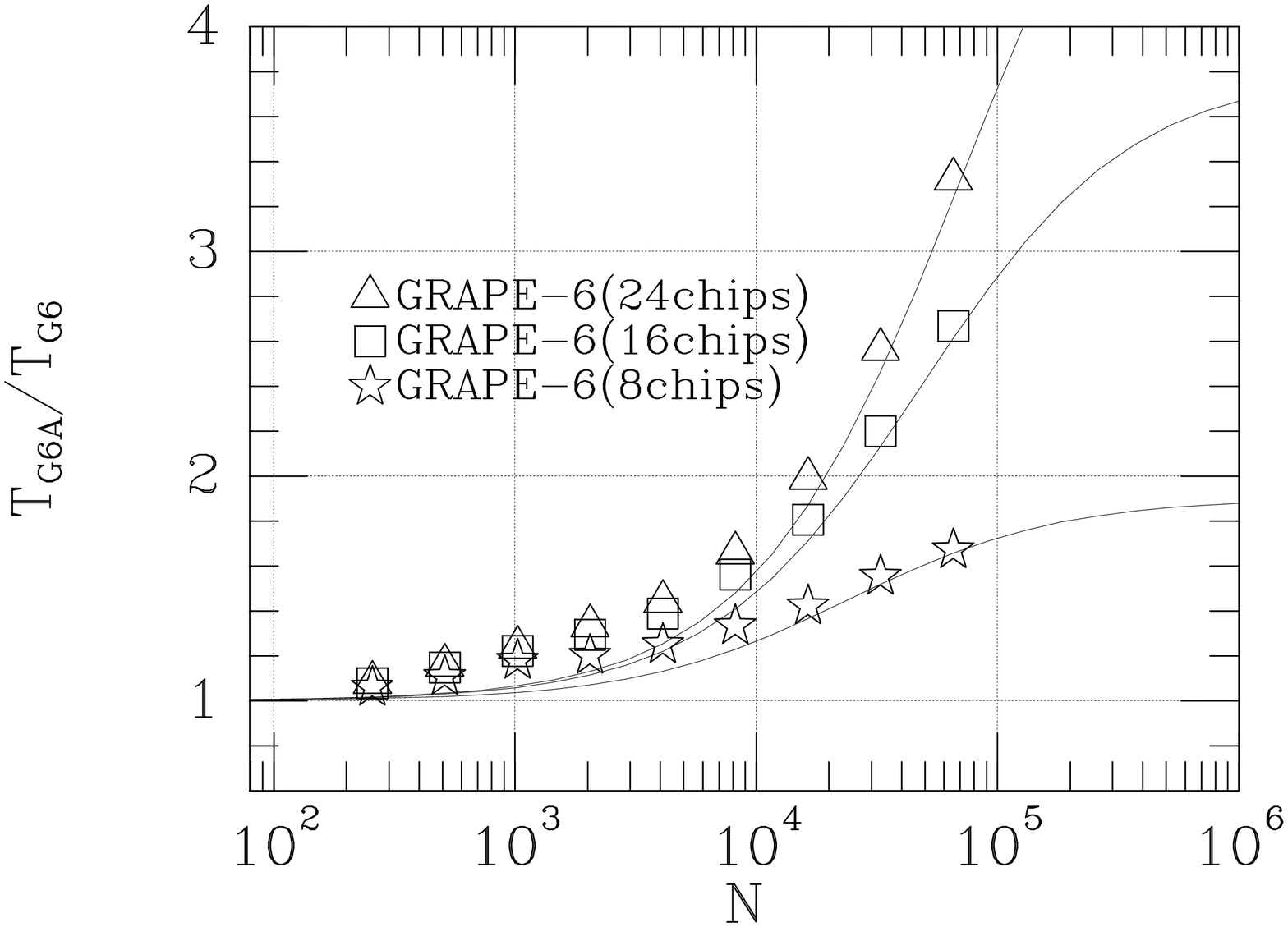}
\end{center}
\caption{Calculation speed of the individual timestep algorithm on
GRAPE-6 processor board (for three different numbers of chips)
relative to that on GRAPE-6A, as a function of the number of particle $N$.  The
solid curves indicate times estimated with the theoretical model in
the text.}
\label{its:fig3}
\end{figure}

Figure \ref{its:fig3} shows the speed of the individual timestep
algorithm on GRAPE-6 processor board (for three different numbers of
chips) relative to that on GRAPE-6A, as a function of the number of
particles $N$. Here, relative speed is defined as $T_{G6A}/T_{G6}$,
where $T_{G6A}$ and $T_{G6}$ are the calculation times for GRAPE-6A
and GRAPE-6, respectively. For large number of particles, GRAPE-6 with
higher theoretical peak speed offers a better performance. However,
the advantage is  small for $N < 5000$.

\subsubsection{Sophisticated program for realistic star cluster simulation}

\begin{figure}
\begin{center}
\leavevmode
\FigureFile(10 cm,8 cm){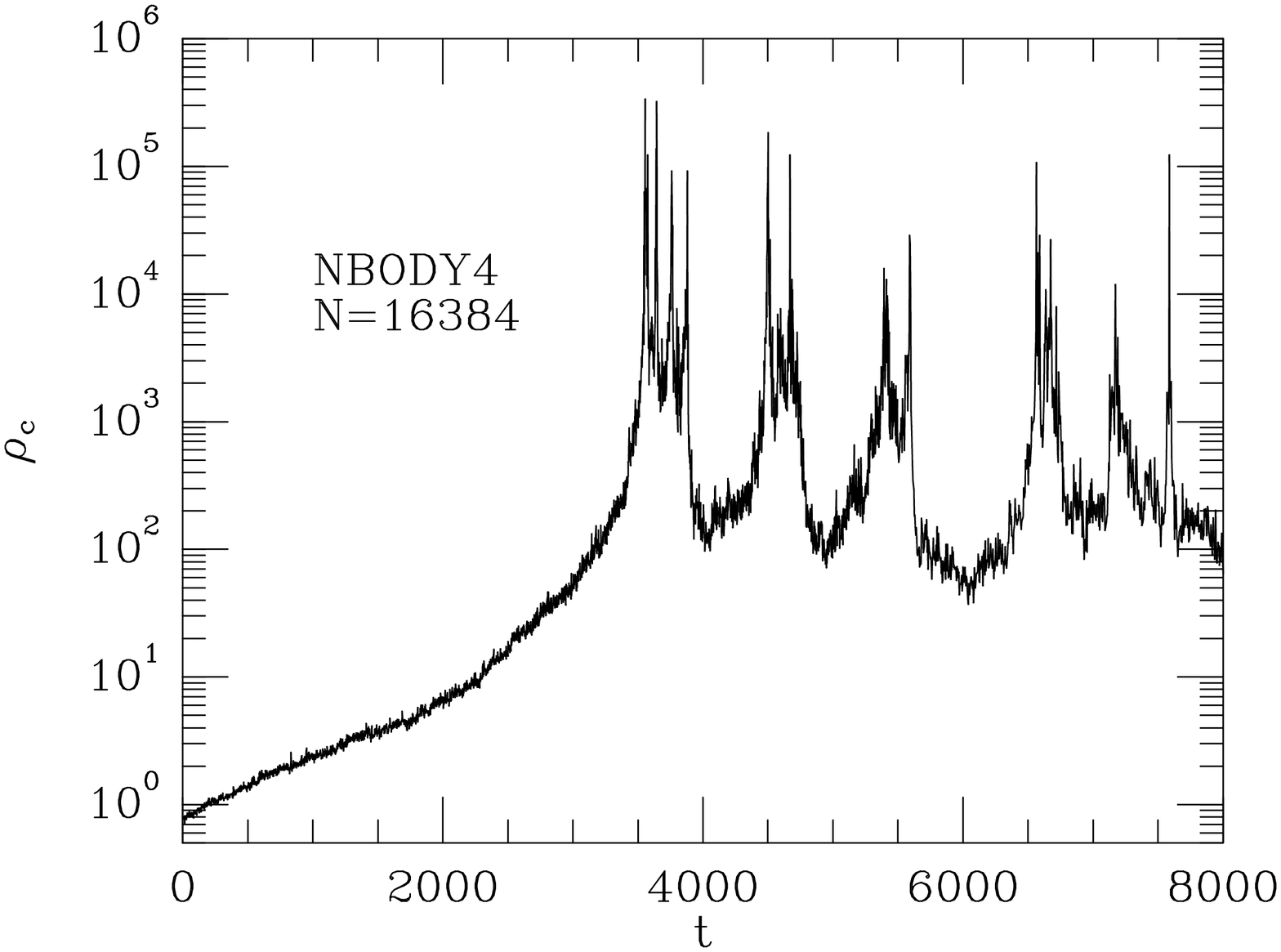}
\end{center}
\caption{Evolution of the core radius for the Plummer model
($N=16384$) computed with {\tt NBODY4} on GRAPE-6A.}
\label{its:fig4}
\end{figure}

Although the individual timestep algorithm is a powerful tool,
simulation of a star cluster requires much more than just the
individual timestep algorithm (see. {\it e.g.}, Aarseth 2003). For
example, special treatments for close encounters and hard binaries are
necessary. One of the programs with all necessary ingredients is {\tt
NBODY4} by Sverre Aarseth, which is developed for GRAPE-4 and GRAPE-6.
Figure \ref{its:fig4} shows a sample result computed with {\tt NBODY4}
on GRAPE-6A. This sample run of $N=16384$ Plummer model completed
within 10 days. The average computing speed was 52.8 Gflops, or around
40 \% of the theoretical peak speed.

%%%%%%%%%%%%%%%%%%%%%%%%%%%%%%%%%%%%%%%%%%%%%%%
\section{Parallel GRAPE-6A cluster System}

We constructed a 24-node PC-GRAPE cluster using GRAPE-6A in University
of Tokyo. Figure \ref{para:fig1} shows the overall structure of the
cluster.  Each node consists of a host computer and one
GRAPE-6A board.  Half of the  host computers have Intel Pentium 4 2.8CGHz
processor (i865G chipset) and 2GB of PC3200 memories and the other
half have  Intel Pentium 4 2.4BGHz proecssors (E7205 chipset) and 2GB of PC2100
memories.  All nodes are connected through 
1000BaseT Ethernet via one single network-switch (Planex
SW-0024G). Each host computer has a network interface card or a
on-board interface with Realtek 8169 single-chip Ethernet controller.
Figure \ref{para:fig2} is a photograph of the cluster.

\begin{figure}
\begin{center}
\leavevmode
\FigureFile(12 cm,10 cm){./g6ac.eps}
\end{center}
\caption{Overall structure of the parallel GRAPE-6A cluster}
\label{para:fig1}
\end{figure}

\begin{figure}
\begin{center}
\leavevmode
\FigureFile(10 cm,8 cm){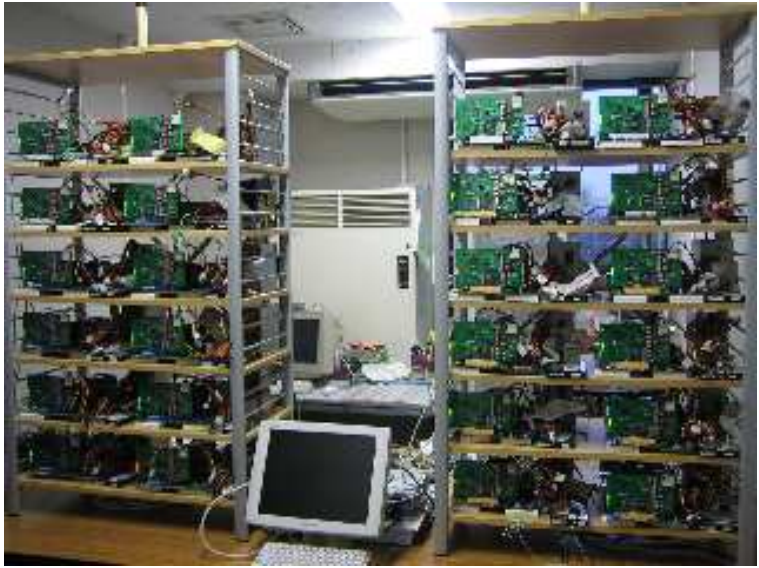}
\end{center}
\caption{Photograph of the parallel GRAPE-6A cluster}
\label{para:fig2}
\end{figure}

This system has the theoretical peak speed of 3.1 Tflops, and is the
first PC cluster system constructed with GRAPE-6A. Similar systems
have been constructed in Rochester, Heidelberg, NAOJ and other places.

%%%%%%%%%%%%%%%%%%%%%%%%%%%%%%%%%%%%%%%%%%%%%%%
\section{Parallel Algorithm}

In this section, we discuss the implementation and performance of
parallel versions of tree algorithm, TreePM algorithm, and individual
timestep algorithm on the parallel GRAPE-6A system. In the performance
measurement in this section, first eight hosts are equipped with
Pentium 4 2.8GHz and the rest with the same processor but with
2.4GHz. For all tests, we use gcc complier (version 3.2.3) and MPICH
(version 1.2.5.2) library on RedHat 9.0.

\subsection{Parallel Tree Algorithm}

At present, there are three parallel implementation of treecode with
GRAPE, each of which one of the authors of this paper (A.K., J.M.,
T.F.) developed.  The main difference between these codes is in the
spatial decomposition scheme.

The parallel code developed by A.K. (hereafter, AK code) uses
essentially the same scheme as Warren and Salmon's (1993) Hashed
Oct-Tree algorithm (with Peano-Hilbert ordering). The performance of
the code was briefly reported in Kawai, Makino (2003). The source code
is available upon request.

\begin{figure}
\begin{center}
\leavevmode
\FigureFile(10 cm,15 cm){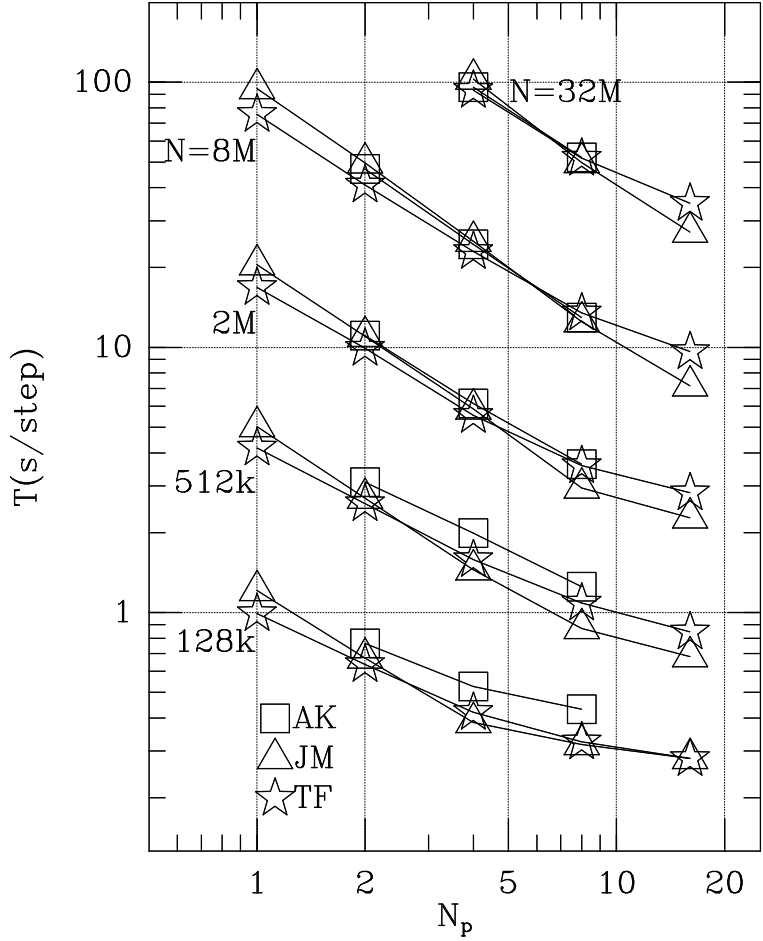}
\end{center}
\caption{Calculation time for the parallel tree algorithms
($\theta=0.75$) as a function of number of node $N_p$. The square,
triangle, star symbols indicate those by AK, JM, and TF codes,
respectively.}
\label{tpara:fig1}
\end{figure}

The parallel code developed by J.M. (hereafter, JM code) uses
orthogonal recursive multi-section, a generalization of the widely
used ORB tree that allows the division to an arbitrary number of domains
in one dimension, instead of allowing only bisection. Details for this
code are described in Makino (2004). The source code is available on a
web site\footnote{\tt
http://grape.s.u-tokyo.ac.jp/pub/people/makino/softwares/pC++tree/}.
Both programs {\tt nbody\_g5} and {\tt nbody\_g6} in this package
operate on GRAPE-6A, and we use {\tt nbody\_g5} to measure the
performance for Figure \ref{tpara:fig1}.

The parallel code  developed by T.F. (hereafter, TF code) uses the
slice (1D) spatial decomposition scheme, similar to the parallel $\rm
P^3M$  coded discussed in Brieu and Evrard (2000). 
The calculation procedure of TF code is summarized as follows:
\begin{itemize}
\item[(1)] Simulation domain is decomposed into the space slices, each
of which is assigned to a node, and particles are distributed into the
nodes. The boundaries of the space slice is determined so that the
numbers of particles in each slice are equal, by the sampling method
(Blackston and Suel 1997).
\item[(2)] Each node constructs a tree structure 
and makes interaction lists for all other nodes.
\item[(3)] Each node receives the interaction lists from all other
nodes, and reconstruct the tree structure which includes all particles
and nodes in the interaction lists it received. 
\item[(4)] Forces on the particles in each node are calculated
using the tree structure.
\end{itemize}

Both TF code and JM code use the interaction list to represent the
information of the tree of one node necessary for the force
calculation of particles in another node. In the original ORB
algorithm (Warren and Salmon 1993), what is called local essential
tree was constructed and exchanged. The use of the interaction list
reduces the complexity of the program and amount of communication, for
the slight increase in the cost of the tree construction. 

Figure \ref{tpara:fig1} shows the calculation time per one timestep as
a function of the number of nodes, $N_p$, for the three parallel codes
described above. For the particle distribution, we used an uniform
sphere of equal mass particles.  We could not measure the calculation
time with $N_p<4$ for $N=$32M because of the limitation of the size of
the main memory of the host computers.  The opening angle was
$\theta=0.75$ and use used $n_{\rm crit}=2000$.  Three parallel codes
show rather similar performances. The JM code exhibits a somewhat
better scalability for large $N_p$.

\subsection{Parallel TreePM Algorithm}

\begin{figure}
\begin{center}
\leavevmode
\FigureFile(8 cm,6 cm){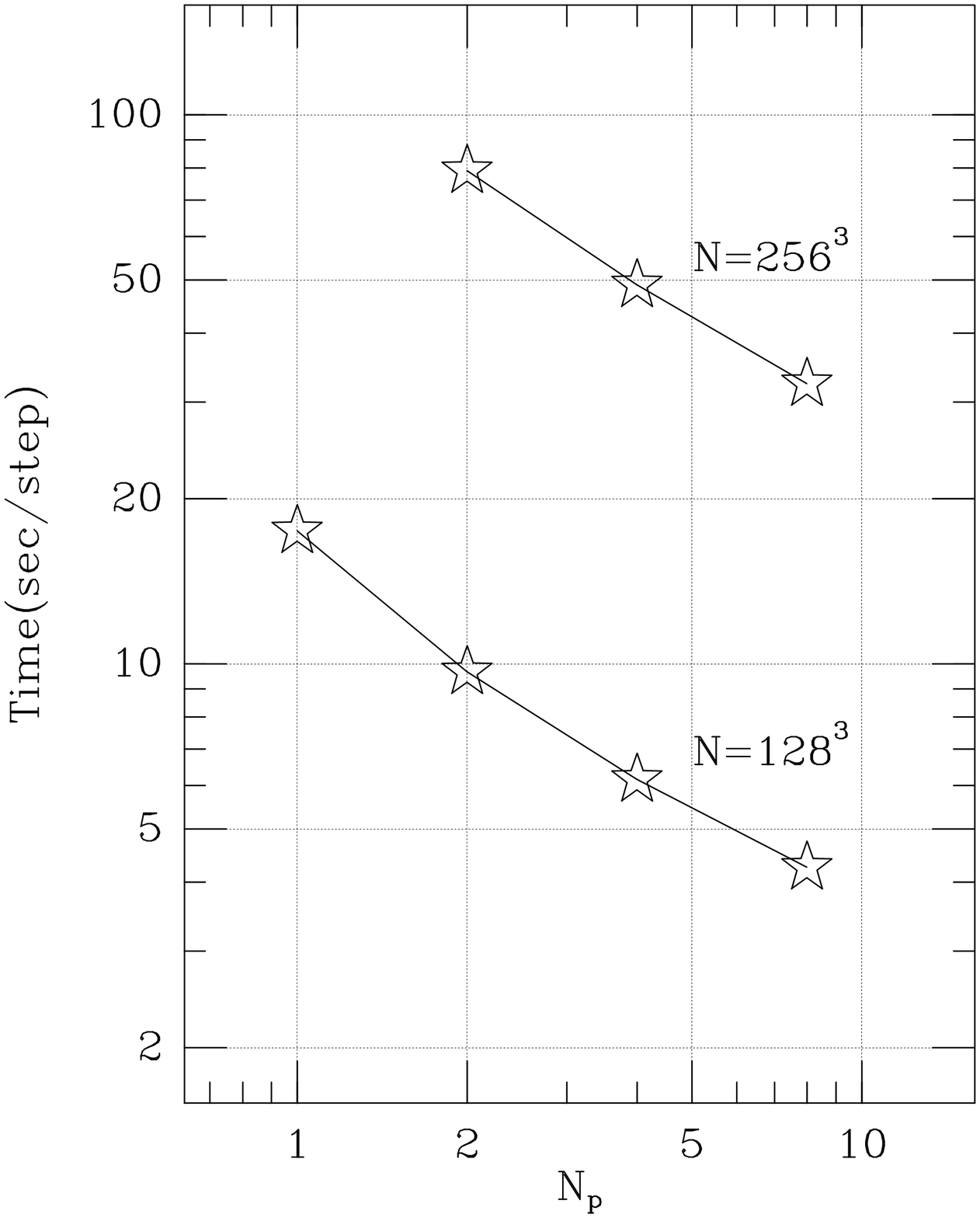}
\end{center}
\caption{Calculation time for the parallel TreePM algorithm as a
function of the number of node $N_p$.}
\label{ppara:fig1}
\end{figure}

The  parallel code for the TreePM algorithm we developed is  a variant of
the TF code discussed in the previous subsection. It uses the
slice (1D) decomposition scheme. Figure \ref{ppara:fig1} shows the
calculation time per timestep as a function of the number of nodes,
$N_p$, at the initial redshift. The particle distribution and other
parameters are the same as in section \ref{sec:treepm}. For more
details, see Yoshikawa and Fukushige (2005).

\subsection{Parallel Individual Timestep Algorithm}
\label{sec:pits}

Here we discuss a parallel implementation of the individual timestep
algorithm for the parallel GRAPE-6A system. This algorithm uses
so-called "$j$-parallelization" (Makino et al 1997), in which force
calculation on a particle is performed in parallel. An important
advantage of this algorithm is that the maximum number of particles
we can use is proportional to the number of GRAPE-6A board.  In the
current implementation, all nodes have complete copy of all particle
data.  The algorithm is summarized as follows:

\begin{itemize}
\item[1.] As the initialization procedure, Each node sends data of
$N/N_p$ particles to the particle memory of  GRAPE-6A.

\item[2.] Each node creates the list of particles to be
integrated at the present timestep. 

\item[3.] Repeat 4.-9. for all particles in the list. 

\item[4.] Each node predicts the position and velocity of the 
particle. 

\item[5.] Each node sends the predicted position and velocity to
GRAPE-6A.

\item[6.] GRAPE-6A calculates partial forces from $N/N_p$ particles,
and sends them back to the host.

\item[7.] The global summation and broadcast of the returned partial
forces is performed, so that all nodes have the same total forces for
all particles in the list. We use MPI {\tt AllReduce()} library function.

\item[8.] Each node integrates the orbits of the particles and
determines new timestep.

\item[9.] Each node sends the updated particle data to the particle
memories on GRAPE-6A if they are already stored in step 1.

\item[10.] Each node updates the present system time and go back to
step 2.

\end{itemize}
In this implementation, only the force calculation is done in
parallel. Each node does the orbit integration for all particles. This
might seem a bit wasteful, but actually this algorithm is close to optimum,
since it minimizes the amount of interprocessor communication. 

\begin{figure}
\begin{center}
\leavevmode
\FigureFile(10 cm,8 cm){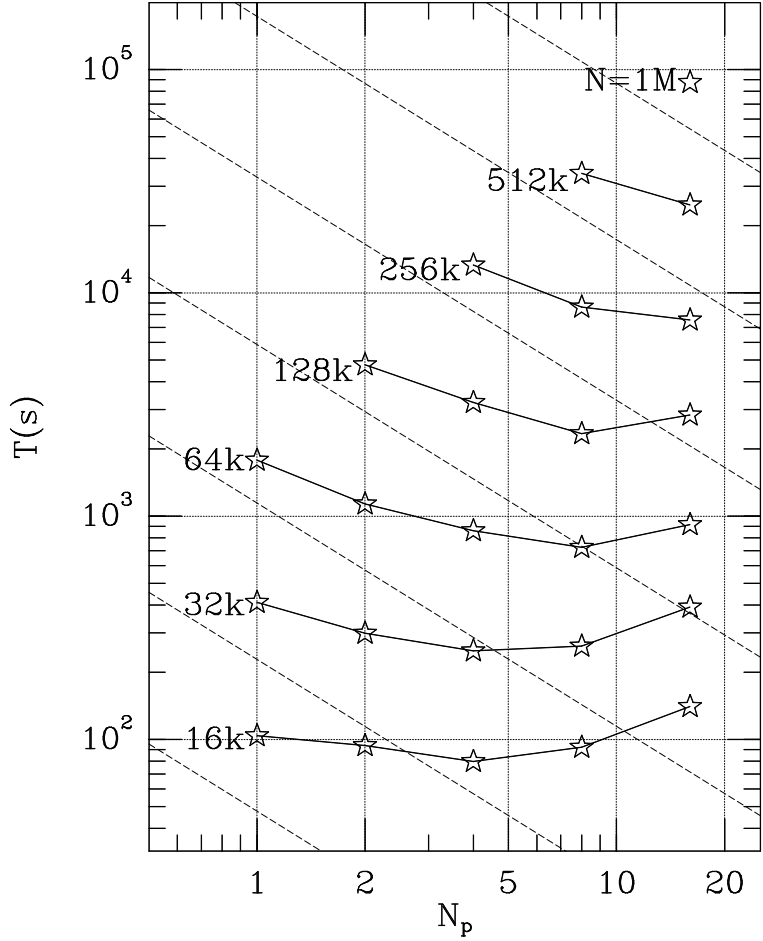}
\end{center}
\caption{Calculation time per one time unit for the parallel
individual timestep algorithm as a function of number of node
$N_p$. Dashed lines indicate times spent for the force calculation
on GRAPE-6A. }
\label{ipara:fig1}
\end{figure}

\begin{figure}
\begin{center}
\leavevmode
\FigureFile(10 cm,8 cm){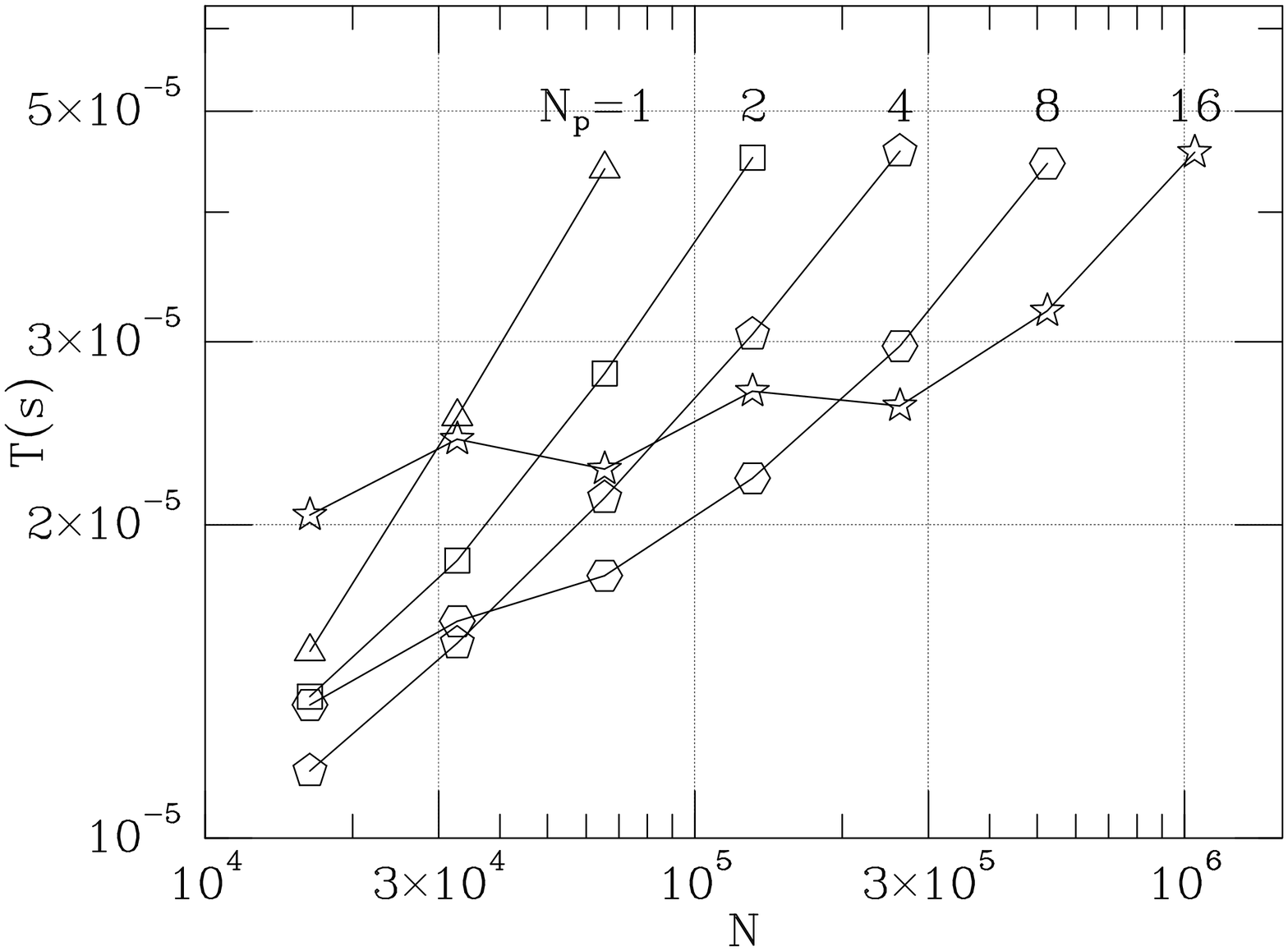}
\end{center}
\caption{Calculation time per one particle step for the parallel
individual timestep algorithm as a function of number of particles 
$N$.}
\label{ipara:fig4}
\end{figure}

Figure \ref{ipara:fig1} shows the calculation time for the integration
of the same system as used in section \ref{sec:its} for one time unit
as a function of the number of node, $N_p$.  We set $\eta=0.01$ and
$\varepsilon=4/N$. We actually measured the calculation time for
integration from time 3/4 to 1.0 for $N\le 32$k and that from 3/32 to
4/32 for $N \ge 64$k, respectively, and scaled the result to give the
calculation time for one time unit. Figure \ref{ipara:fig4} shows the
calculation time per one particle step as a function of number of
particles, $N$.

\begin{figure}
\begin{center}
\leavevmode
\FigureFile(10 cm,8 cm){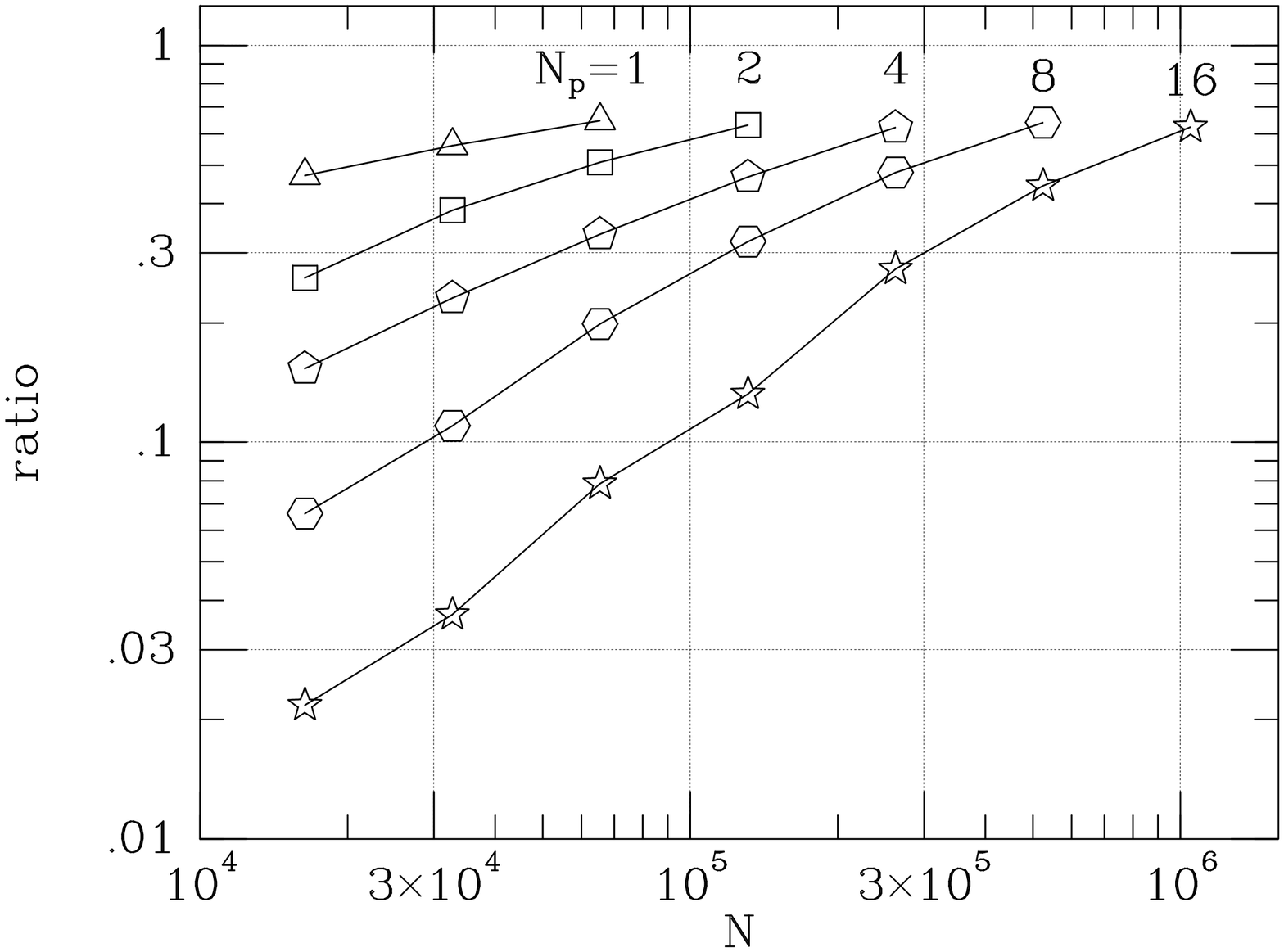}
\end{center}
\caption{Ratio of the sustained calculation speed to the peak speed
for the parallel individual timestep algorithm as a function of $N$.}
\label{ipara:fig2}
\end{figure}

\begin{table}
\caption{Breakdown of the calculation time per one particle step for
the parallel individual timestep algorithm}
\begin{center}
\begin{tabular}{cc|cccc}
\hline
\hline
$N$ & $N_p$ & $t$(s) & $t_{\rm gr}$(s) & $t_{\rm host}$(s) & $t_{\rm MPI}$(s) \\
\hline
65536 & 1 & $4.39 \times 10^{-5}$ & $4.23 \times 10^{-5}$  & $1.60\times 10^{-6}$ & $0$ \\
65536 & 2 & $2.79 \times 10^{-5}$ & $2.27 \times 10^{-5}$  & $1.57\times 10^{-6}$ & $3.57\times 10^{-6}$ \\
65536 & 4 & $2.12 \times 10^{-5}$ & $1.29 \times 10^{-5}$  & $1.60\times 10^{-6}$ & $6.66\times 10^{-6}$ \\
65536 & 8 & $1.78 \times 10^{-5}$ & $8.06 \times 10^{-6}$  & $1.60\times 10^{-6}$ & $8.18\times 10^{-6}$ \\
65536 & 16 & $2.26 \times 10^{-5}$ & $5.59 \times 10^{-6}$  & $1.62\times 10^{-6}$ & $1.53\times 10^{-5}$ \\
\hline
262144 & 4 & $4.57 \times 10^{-5}$ & $3.69 \times 10^{-5}$  & $1.84\times 10^{-6}$ & $6.91\times 10^{-6}$ \\
262144 & 8 & $2.97 \times 10^{-5}$ & $2.00 \times 10^{-5}$  & $1.87\times 10^{-6}$ & $7.75\times 10^{-6}$ \\
262144 & 16 & $2.60 \times 10^{-5}$ & $1.16 \times 10^{-5}$  & $1.87\times 10^{-6}$ & $1.25\times 10^{-5}$ \\
\hline
1048576 & 16 & $4.56 \times 10^{-5}$ & $3.39 \times 10^{-5}$ & $2.25\times 10^{-6}$ & $9.42\times 10^{-6}$ \\
\hline
\end{tabular}
\end{center}
\label{ipara:tab1}
\end{table}

We can see that for large $N$, our parallel code achieves good
efficiency. However, for small values of $N$ such as 16,384, the
speedup is rather marginal and the parallel calculation with $N_p= 16$
is slightly slower than the calculation with single GRAPE-6A. In other
words, parallel scalability is not very good.  This is mainly because
of the relatively short message length for the internode
communication. Such short message is inevitable with the individual
timestep algorithm. The relatively long message latency of the MPICH
implementation we used resulted in this rather poor scaling
characteristic.  Figure \ref{ipara:fig2} shows the ratio of the
sustained calculation speed to the peak speed as a function of $N$.
Table \ref{ipara:tab1} shows the breakdown of the calculation times
per one particle step for some representative $N$ and $N_p$. Here,
$t_{\rm gr}$, $t_{\rm host}$ and $t_{\rm grape}$ indicate the time
spent on GRAPE-6A and for communication between host computer and
GRAPE-6A (step 1, 5, 6, 9), the time spent on the host computer (step
2, 4, 8, 10), and the time spent for the internode communication (step
7).

\begin{figure}
\begin{center}
\leavevmode
\FigureFile(10 cm,8 cm){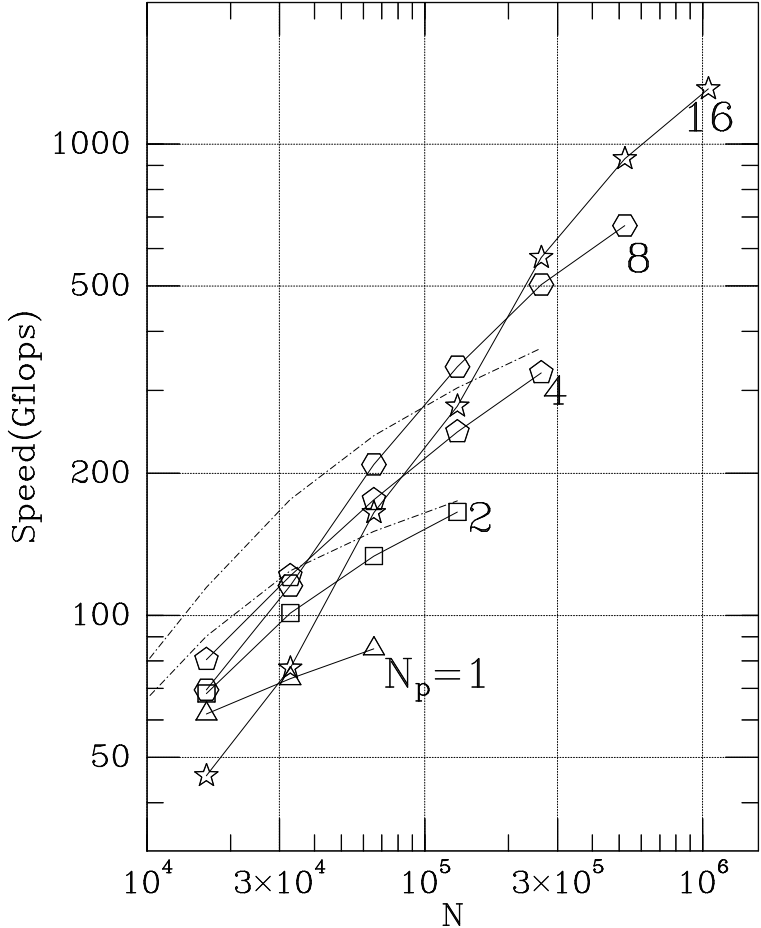}
\end{center}
\caption{Calculation speed in Gflops for the parallel individual
timestep algorithm as a function of number of particles $N$. The
dot-dashed curves indicate that for GRAPE-6 with 16 chips (upper) and
8 chips (lower).}
\label{ipara:fig3}
\end{figure}

Figure \ref{ipara:fig2} shows the calculation speed in Gflops as a
function of $N$. We also plot the performance of GRAPE-6 with 16
chips(4 modules) and 8 chips(2 modules). The performance of the
parallel GRAPE-6A system is close to that of GRAPE-6 with the same
number of modules for $N\ge 64$k.

%%%%%%%%%%%%%%%%%%%%%%%%%%%%%%%%%%%%%%%%%%%%%%%
\section{Summary and Discussion}

We have developed a special-purpose computer for astrophysical
$N$-body simulations, GRAPE-6A, which is a downsized version of
GRAPE-6 suitable for parallel PC cluster configurations. With
GRAPE-6A, it becomes practical to construct larger-scale PC-GRAPE 
cluster systems. Various  parallel implementation of important
algorithms  such as treecode and individual timestep are already
running  on our parallel GRAPE-6A system with very good performance.
For example, using the tree algorithm on our parallel system, we can
complete a collisionless simulation with 100 million particles (8000
steps) within 10 days.

Several institutes constructed similar parallel GRAPE-6A
clusters. Tsukuba University started the project to construct an
impressive 256-node cluster for the simulation of first-generation
objects (Umemura et al. {\tt http://www.rccp.tsukuba.ac.jp}). 

At present, GRAPE-6A is probably the best solution for construction of
PC-GRAPE cluster.  Further improvement of GRAPE-6A with up-to-date
technology would provide even more powerful computing systems. As is
clear in the breakdown in Table \ref{per:tab1} and \ref{its:tab1}, the
total performance is limited by the communication speed between the
host computer and GRAPE.  For the communication, GRAPE-6A uses PCI
(32bit/33MHz) interface, which is rather old technology.  Faster
interfaces, such as PCI-X or PCI Express, are now available. We are
currently developing a new version of GRAPE with PCI-X interface. The
peak speed of PCI-X (64bit/133MHz) is 1.06GByte/s. This speed would be
fast enough compared to the speed of the host computer for several
years to come.

%%%%%%%%%%%%%%%%%%%%%%%%%%%%%%%%%%%%%%%%%%%%%%%
\section*{Acknowledgments}

We are grateful to Hiroshi Daisaka and Eiichiro Kokubo for discussions
on the implementation on various host computers and to Kohji Yoshikawa
for helping with implementation of the TreePM algorithm.  We would
like to thank all of those who involved in the GRAPE project. This
research was supported by the Research for the Future Program of Japan
Society for the Promotion of Science (JSPS-RFTF97P01102), the
Grants-in-Aid by the Japan Society for the Promotion of Science
(14740127) and by the Ministry of Education, Science, Sports, and
Culture of Japan (16684002).

\end{document}